\def\e{{\rm e}}
\def\del{\partial}
\def\half{{1\over2}}

\def\abs#1{{\left|{#1}\right|}}
\def\vev#1{\langle #1 \rangle}
\newcommand{\PRD}[3]{Phys. Rev. {\bf D{#1}} (19{#2}) {#3}} 
 
\newcommand{\NPB}[3]{Nucl. Phys. {\bf B{#1}} (19{#2}) {#3}} 
\newcommand{\PLB}[3]{Phys. Lett. {\bf B{#1}} (19{#2}) {#3}} 
 
\newcommand{\ANN}[3]{Ann. Phys. {\bf {#1}} (19{#2}) {#3}} 
 
\newcommand{\RPP}[3]{Rep. Prog. Phys. {\bf {#1}} (19{#2}) {#3}} 
 
\newcommand{\hmu}{\hat\mu}
\newcommand{\hnu}{\hat\nu}
\newcommand{\hrho}{\hat\rho}
\newcommand{\hsigma}{\hat\sigma}
\documentstyle[12pt,epsf]{article}
\begin{document}
\title{Softly Broken Supersymmetric Gauge Theories through 
Compactifications
}
\author{Kazunori Takenaga $^{}$\thanks {email: 
takenaga@oct.phys.kobe-u.ac.jp,~~JSPS Research Fellow}\\ \it {Department 
of Physics, Kobe University,}\\ {\it Rokkodai, Nada, Kobe 657, Japan}
}
\date{} 
\maketitle
\baselineskip=18pt
\vskip 3cm
\begin{abstract}
Effects of boundary
conditions of fields for compactified space directions on the 
supersymmetric gauge theories are discussed. 
For general and possible boundary conditions the supersymmetry is 
explicitly broken to yield universal soft supersymmetry breaking 
terms, and the gauge symmetry of the theory can also be broken 
through the dynamics of non-integrable phases, depending on number 
and the representation under the gauge group of matters.
The $4$-dimensional supersymmetric QCD is studied as a toy model
when one of the space coordinates is compactified on $S^1$. 
\end{abstract}
\vskip 4cm
\addtolength{\parindent}{2pt}
\newpage
\section{Introduction}\qquad
Our space-time dimensions may be larger than four at fundamental 
scale such as the Planck scale. Actually, it must be so for 
consistency of the (super) string theories\cite{gsw}. 
In our laboratories, however, we know that our space-time dimensions are 
four, so that extra space coordinates must be compactified by certain 
mechanism. Mechanism of the compactifications is still unknown, but 
physical consequences of 
compactifications have been studied in various 
theories since the proposal by Kaluza-Klain\cite{kaluza}. 
\par
One must specify boundary conditions of fields for 
compactified directions if space is multiply-connected. 
We do not know, {\it a priori}, what boundary conditions should be imposed 
on the fields for the directions. This is very contrary to the case of the 
finite temperature field theory, in which the boundary conditions for the 
euclidean time direction is determined definitely by the quantum 
statistics of particles.\par
We shall consider general and possible boundary conditions 
in supersymmetric gauge theory.
One can require that the fields return to their original values up to phases 
proportional to their charges of global 
symmetry transformations when the fields travel along the 
compactified directions\cite{bhosotani}\cite{ahosotani}.
The global symmetry transformations must be symmetry of the theory.
The lagrangian is automatically single-valued even if the fields have 
such the boundary conditions. \par
In a previous paper\cite{takenaga} we studied the effect of 
the boundary condition 
associated with the $U(1)_{R}$ symmetry on the supersymmetry breaking
in the supersymmetric QED.
The translational invariance for the compactified direction is broken 
by the boundary condition, so that the variation of action under the 
supersymmetric transformations 
does not vanish and remains as surface terms. 
The supersymmetry is explicitly broken due to the 
boundary condition. 
All the effects of the supersymmetry breaking turn out 
to appear in the lagrangian as the soft supersymmetry breaking terms 
whose coupling constants are given by an unique 
parameter and the gauge coupling. 
\par
In addition to the boundary condition mentioned above, we can consider 
the boundary condition associated with the global gauge symmetry in 
supersymmetric gauge theories. The boundary condition is closely 
related with the non-integrable phases of the gauge field along 
the compactified direction, which is dynamical degrees of freedom in 
multiply-connected space\cite{bhosotani}\cite{ahosotani}. 
The boundary condition does not 
break the supersymmetry, but instead, it can break the gauge symmetry 
of the theory through the dynamics of the non-integrable phases.
\par 
In this paper we shall investigate the supersymmetric 
gauge theory, namely, the supersymmetric QCD (SQCD) 
when the theory, in which the fields have general and 
possible boundary conditions for the compactified 
directions, is compactified.
In order to study it as analytically as possible, we shall consider a 
toy model such that the $SU(N)$ SQCD is compactified on $M^3\otimes S^1$.
In the next section we will define the boundary conditions of the fields 
for the $S^1$ direction. And we will discuss how the boundary 
condition associated with the $U(1)_{R}$ symmetry breaks the 
supersymmetry and how the soft supersymmetry breaking terms appear in 
the lagrangian. In the section $3$ we will give brief summary on the 
dynamics of the non-integrable phases in multiply-connected space. 
Then, we 
will evaluate the effective potential for the non-integrable phases
in the SQCD to find how the gauge symmetry is 
broken. The symmetry breaking depends on number and the 
representation under the gauge group of matters. The final section is 
devoted to conclusions and discussions.\par
\section{Soft Supersymmetry Breaking Terms}
In this section we shall show how the boundary condition associated with 
the $U(1)_{R}$ symmetry breaks the supersymmetry and how the soft 
supersymmetry breaking terms appear in a toy model, $4$-dimensional 
$SU(N)$ SQCD compactified on $M^3\otimes S^1$, where $M^3$ is the 
$3$-dimensional Minkowski space-time, and $S^1$ is a circle. 
We use a notation 
such as $x^{\hmu}\equiv (x^{\mu}, x^3)\equiv (x, y)$ and
denote the length of the circumference of the $S^1$ by $L$.
\par
\subsection{Boundary Conditions and Surface Terms}
The lagrangian we consider is given by
\begin{eqnarray}
{\cal L}_{SQCD}&=&
{1\over 2}{\rm tr}\Bigl[W^AW_{A} + h. c.\Bigr]_{F}\nonumber \\
&=&{\rm tr}\Bigl[
-\half F_{\hmu\hnu}F^{\hmu\hnu}
-i\lambda\sigma^{\hmu}D_{\hmu}{\bar\lambda} 
+iD_{\hmu}\lambda\sigma^{\hmu}{\bar\lambda}+D^2 \Bigr].
\label{eq:lagrangian}
\end{eqnarray}
The subscript $F$ in (\ref{eq:lagrangian}) means $F$-term. The $W_{A}$ is 
the spinorial chiral superfield constructed by the vector 
superfield $V$ in the Wess-Zumino gauge\cite{wessb}. 
The $A(=1, 2)$ stands for a two-component Weyl spinor index. 
The $W_{A}(\equiv W_{A}^aT^a)$ contains the vector boson 
$A_{\hmu}$ (gluon), a two-component Weyl fermion $\lambda_{A}$ (gaugino) 
and the auxiliary field $D$. The auxiliary 
field is eliminated by the equations of motion for it.
The $T^a(a=1,\cdots,N^2-1)$ is the generator of $SU(N)$ gauge group. 
The $F_{\hmu\hnu}$ is the field strength for the gluon.
\par
Under the supersymmetric transformations defined by
$$
\begin{array}{ll}
\delta_{\xi}A_{\hmu}
=\xi \sigma_{\hmu}{\bar\lambda}-{\bar\xi}{\bar\sigma}_{\hmu}\lambda,& 
\delta_{\xi}\lambda =\xi D+\sigma^{\hmu\hnu}\xi F_{\hmu\hnu}, \\[2mm]
\delta_{\xi}D =i\xi \sigma^{\hmu}D_{\hmu}{\bar \lambda} + 
i{\bar\xi}{\bar\sigma}^{\hmu}D_{\hmu}\lambda,&\\[2mm] 
\end{array}
$$
the lagrangian varies as
$\delta_{\xi}{\cal L}_{SQCD}=\del_{\hmu}X^{\hmu}$, where $X^{\hmu}$ is 
calculated as
$$
X^{\hmu}=\Bigl[-\xi \sigma_{\hnu}{\bar\lambda}^aF^{a\hmu\hnu} +{i\over 
2}\xi \sigma^{\hrho\hsigma}{\sigma}^{\hmu} 
{\bar\lambda}^aF_{\hrho\hsigma}^a
+{i\over 2}\xi \sigma^{\hmu}{\bar\lambda}^aD^a+h.c.\Bigr].
$$
The $\xi$ is the supersymmetric transformation parameter of a 
two-component constant Weyl spinor. 
The lagrangian (\ref{eq:lagrangian}) is invariant under the $U(1)_{R}$ 
transformation defined by
$$
W_{A}(\theta)\rightarrow {\rm e}^{i\beta}W_{A}({\rm e}^{-i\beta}\theta).
$$
The $\theta$ is the superspace coordinate. In terms of the component 
fields, the transformations can be written as 
$$
\begin{array}{lll}
\lambda_{A} \rightarrow {\rm e}^{i\beta}\lambda_{A}, & A_{\hmu}\rightarrow 
A_{\hmu},& D\rightarrow D,\\ [2mm] 
\end{array}
$$
We see that the $U(1)_{R}$ charges are different between the bosons and 
the fermions in a supermultiplet $W_{A}$. 
\par 
We define the 
boundary conditions of the fields for the $S^1$-direction as follows;
\begin{eqnarray}
A_{\hmu}(x, y+L)&=&U_{g}A_{\hmu}(x, y)U_{g}^{\dagger},\nonumber \\
\lambda(x, y+L)&=&{\rm e}^{i\beta}U_{g}\lambda(x, y)U_{g}^{\dagger},
\nonumber \\
D(x, y+L)&=&U_{g}D(x, y)U_{g}^{\dagger},
\label{eq:boundarya}
\end{eqnarray}
where $U_{g}$ is a constant $SU(N)$ matrix.
The lagrangian (\ref{eq:lagrangian}) is still single-valued even if the 
fields have such the boundary conditions.
If the fields have the boundary condition (\ref{eq:boundarya}), 
the surface term, 
$$
\delta_{\xi}S_{SQCD}=\Bigl(\int dx~ X^{3}(x, y)\Bigr) \vert_{S^1}
$$
does not vanish because there is a difference between $X^3(x, y+L)$ and 
$X^3(x, y)$ due to the non-trivial phase ${\rm e}^{i\beta}$ 
in (\ref{eq:boundarya}). Note that the supersymmetric 
transformation parameter $\xi$ obeys the periodic boundary condition.  
The $X^3$ is the third space-component of the total 
derivative $X^{\hmu}$. The translational invariance for 
the $S^1$ direction is broken by the boundary condition, so that the 
supersymmetry is explicitly broken.
Note that $U_{g}$ associated with the global gauge symmetry does not 
break the translational or supersymmetric invariance.
\par 
\subsection{Gauged $U(1)_{R}$ Transformation}
For a moment, let us notice the boundary condition associated with 
the $U(1)_{R}$ symmetry, that is, the non-trivial phase
${\rm e}^{i\beta}$ in (\ref{eq:boundarya}). 
We shall discuss the importance of $U_{g}$ on the gauge symmetry 
breaking in the section $3$.
\par
When we expand the fields in the Fourier series for the $S^1$ direction,
\begin{eqnarray}
\begin{array}{ll}
A_{\hmu}(x, y) = {1\over{\sqrt L}}\sum_{n=-\infty}^{+\infty} 
A_{\hmu}^{(n)}(x)\e^{{{2\pi i}\over L}ny},& 
\lambda(x, y)={1\over{\sqrt 
L}}\sum_{n=-\infty}^{+\infty} \lambda^{(n)}(x)\e^{{{2\pi i}\over 
L}(n+{{\beta} \over {2\pi}})y}, \\ [2mm]
D(x, y)={1\over{\sqrt L}}\sum_{n=-\infty}^{+\infty} D^{(n)}(x)\e^{{{2\pi 
i}\over L}ny},&\\[2mm]
\label{eq:fse}
\end{array}
\end{eqnarray}
we observe that $\lambda(x, y)$ given in (\ref{eq:fse}) 
can be redefined so as to satisfy the periodic boundary condition by 
gauged $U(1)_{R}$ transformation whose parameter depends linearly 
only on the compactified coordinate $y$ \cite{schwartz}\cite{bachas};
\begin{eqnarray}
\lambda (x, y) = U_{R}(y){\tilde{\lambda}}(x, y),\qquad 
U(y)_{R}\equiv {\rm e}^{i{\beta \over L}y},
\label{eq:relation}
\end{eqnarray}
where $\tilde\lambda(x, y)$ satisfies
${\tilde{\lambda}}(x, y+L)={\tilde{\lambda}}(x, y)$. 
\par
As discussed in the previous paper\cite{takenaga}, the supersymmetry 
breaking terms 
become manifest in the lagrangian by redefining the fields so as 
to satisfy the periodic boundary condition.
By using (\ref{eq:relation}), ${\cal L}_{SQCD}$ can be 
recast in terms of $(A_{\hmu}, \tilde\lambda)$ as
$$
{\cal L}_{SQCD}={\tilde{\cal L}}_{SQCD}+ {\tilde {\cal L}}_{SQCD}^{soft},
$$
where ${\tilde{\cal L}}_{SQCD}$ has the same form with the original 
lagrangian except that all the fields satisfy the periodic boundary 
condition. And ${\tilde {\cal L}}_{SQCD}^{soft}$ is obtained as 
$$
{\tilde {\cal L}}_{SQCD}^{soft}=U_{R}\del_{\hmu}U_{R}^{\dagger}
{\rm tr}\Bigl[-2i\tilde\lambda\sigma^{\hmu}\tilde{\bar\lambda}\Bigr]
=-{{2\beta}\over L}{\rm tr}(\bar\psi_{1}\psi_{1}+\bar\psi_{2}\psi_{2}),
$$
where we have used the Majorana
spinors in the $3$-dimensions defined by ${\tilde\lambda_{M}}^T\equiv 
(\psi_{1}, i\psi_{2})^T$.
The ${\tilde\lambda}_{M}$ is the $4$-component Majorana spinors constructed by 
${\tilde\lambda_{M}}^T=
({\tilde\lambda},{\tilde{\bar\lambda}})^T$.
The supersymmetry breaking terms are the gaugino masses whose coupling
constants are given by an unique parameter $\beta$.
The ${\tilde {\cal L}}_{SQCD}^{soft}$ is generated through the 
derivative in the kinetic term for the gaugino, where the
gauged $U(1)_{R}$ transformation (\ref{eq:relation}) is not 
respected as symmetry of the theory. 
\par
We can define the modified supersymmetry transformations 
for $(A_{\hmu}, \tilde\lambda)$. The explicit breaking of 
the supersymmetry due to (\ref{eq:boundarya}) also becomes manifest by 
the variation of ${\cal L}_{SQCD}$ under the modified supersymmetry 
transformations. We find 
\begin{eqnarray}
{\tilde\delta_{\xi}}{\cal L}_{SQCD}&=&
\del_{\hmu}{\tilde X}^{\hmu}+
\bigl[U_{R}\del_{\hmu}U_{R}^{\dagger}\bigr]{\tilde\delta}_{\xi}{\rm tr}
\Bigl[-2i{\tilde\lambda}\sigma^{\hmu}{\tilde{\bar\lambda}} 
\Bigr],
\label{eq:modify}
\end{eqnarray}
where $\tilde\delta_{\xi}$ defines the modified supersymmetric 
transformations. 
The boundary condition associated with the $U(1)_{R}$ symmetry breaks the 
supersymmetry explicitly as shown in the second term in 
(\ref{eq:modify}).
As we expected, the breaking of the supersymmetry is entirely due to the 
locality of the gauged $U(1)_{R}$
transformation,~{\it i.e.} $U_{R}\del_{\hmu}U_{R}^{\dagger}$.
\par
\subsection{Supersymmetry Breaking Terms from Matters}
Let us discuss what types of the supersymmetry breaking 
terms appear if we add matters. 
We introduce massive matter superfields $Q_{I}({\overline Q}^I)$ which 
belong to the (anti) fundamental representation under $SU(N)$ gauge group.
The $I(=1,\cdots, N_{F})$ stands for flavour index. 
The $Q_{I}({\overline Q}^I)$ contains a complex scalar 
field $\phi_{qI}(\phi_{\overline q}^I)$, a two-component Weyl 
fermion $q_{I}({\overline q}^I)$ and the auxiliary 
field $F_{qI}(F_{\overline q}^I)$.
In appendix we present the explicit form of the lagrangian for the 
matters and 
its variation under the supersymmetry transformations in the 
$4$-dimensions.
\par
The $U(1)_{R}$ symmetry, which is symmetry of the theory, is defined by 
\begin{equation}
Q_{I}(\theta)\rightarrow {\rm e}^{i\beta}Q_{I}({\rm e}^{-i\beta}\theta),\qquad
{\overline Q}^I(\theta)\rightarrow 
{\rm e}^{i\beta}{\overline Q}^I({\rm e}^{-i\beta}\theta).
\label{eq:ursym}
\end{equation}
In terms of the component fields, the transformations can be written as
\begin{equation}
\begin{array}{lll}
\phi_{qI} \rightarrow {\rm e}^{i\beta}\phi_{qI},& q_{I}\rightarrow q_{I},
& F_{qI}\rightarrow {\rm e}^{-i\beta}F_{qI},\\ [2mm] 
\phi_{\overline q}^I\rightarrow {\rm e}^{i\beta}\phi_{\overline q}^I,
& {\overline q}^I\rightarrow {\overline q}^I,&
F_{\overline q}^I\rightarrow {\rm e}^{-i\beta}F_{\overline q}^I.\\[2mm]
\end{array}
\label{eq:compsym}
\end{equation}
We see that the $U(1)_{R}$ charges are different between the bosons 
and the fermions in each supermultiplet $Q_{I}, {\overline Q}^{I}$. 
\par
The boundary conditions we take 
are
\begin{equation}
\begin{array}{ll}
\phi_{qI}(x, y+L)={\rm e}^{i\beta}U_{g}\phi_{qI}(x, y), &
\phi_{\overline q}^I(x, y+L)=
{\rm e}^{i\beta}{\overline U}_{g}\phi_{\overline q}^I(x, y),\\[2mm]
q_{I}(x, y+L)=U_{g}q_{I}(x, y),&
{\overline q}^I(x, y+L)={\overline U}_{g}{\overline q}^I(x, y),\\[2mm]
F_{qI}(x, y+L)={\rm e}^{-i\beta}U_{g}F_{qI}(x, y),&
F_{\overline q}^I(x, y+L)=
{\rm e}^{-i\beta}{\overline U}_{g}F_{\overline q}^I(x, y),\\[2mm]
\end{array}
\label{eq:flbc}
\end{equation}
where $U_{g}, {\overline U}_{g}$ are 
constant $SU(N), SU({\overline N})$ matrices, respectively. Their 
importance on the gauge symmetry breaking are discussed in the next section.
Under the supersymmetry transformations, the lagrangian varies as
$\delta_{\xi}{\cal L}_{matters}=\del_{\hmu}X^{\hmu}_{matters}$.
The third space component of the total derivative $X^{3}_{matters}$ 
do not return to their original values after the translation along the 
$S^1$-direction due to the non-trivial phase in (\ref{eq:flbc}). 
Therefore, the supersymmetry is explicitly broken.
\par
When we expand the fields in the Fourier series for 
the $S^1$-direction, we find 
\begin{equation}
\begin{array}{ll}
\phi_{qI}(x, y)=U_{R}(y){\tilde\phi}_{qI}(x, y),&
\phi_{\overline q}^I(x, y)
=U_{R}(y){\tilde\phi}_{\overline q}^I(x, y),\\[2mm]
F_{qI}(x, y)=U_{R}^{\dagger}(y){\tilde F}_{qI}(x, y), &
F_{\overline q}^I(x, y)
=U_{R}^{\dagger}(y){\tilde F}_{\overline q}^I(x, y).\\[2mm]
\end{array}
\label{eq:relationb}
\end{equation}
As before, the fields can be redefined so as to satisfy the periodic 
boundary condition by the gauged $U(1)_{R}$ transformation 
$U_{R}(y)={\rm e}^{i{\beta\over L}y}$.
The ${\tilde\phi}_{qI}, {\tilde\phi}_{\overline q}^I, {\tilde F}_{qI}$ 
and ${\tilde F}_{\overline q}^I$ satisfy the periodic boundary condition.
\par
The supersymmetry breaking terms manifestly appear by 
rewriting the lagrangian in terms of the fields with the 
periodic boundary condition. By using (\ref{eq:relationb}), we obtain
$$
{\cal L}_{SQCD}+{\cal L}_{matters}={\tilde{\cal L}}_{SQCD}+
{\tilde {\cal L}}_{matters}+{\tilde{\cal L}}^{soft},
$$
where ${\tilde {\cal L}}_{SQCD}$ and ${\tilde {\cal L}}_{matters}$
are the same form with the original 
lagrangian except that all the fields satisfy the periodic boundary 
condition. The ${\tilde{\cal L}}^{soft}$ is generated 
through the derivatives in the kinetic terms for the gaugino and 
the squark, where the 
gauged $U(1)_{R}$ transformation by $U_{R}(y)$ is not respected as symmetry 
of the theory. 
The supersymmetry breaking terms are never generated from 
the superpotential $W(Q, {\overline Q})$ because 
there are no derivatives in it. 
The ${\tilde {\cal L}}^{soft}$ is obtained as
\begin{eqnarray}
{\tilde {\cal L}}^{soft}
&=&U_{R}(y)\del_{\hmu}U^{\dagger}_{R}(y) \Bigl[
{\rm tr}\Bigl(-2i{\tilde\lambda}\sigma^{\hmu}{\tilde{\bar\lambda}}\Bigr) 
+\sum_{I}
\Bigl({\tilde \phi}_{qI}^{\dagger}\bigl(D^{\hmu}{\tilde \phi}_{qI}\bigr) 
-\bigl(D^{\hmu}{\tilde \phi}_{qI}\bigr)^{\dagger}{\tilde \phi}_{qI}\Bigr)
\nonumber\\ 
&+&\sum_{I}
\Bigl({\tilde \phi}_{{\overline q}}^{I\dagger}
\bigl(D^{\hmu}{\tilde \phi}_{{\overline q}}^I\bigr) 
-\bigl(D^{\hmu}{\tilde \phi}_{{\overline q}}^I\bigr)^{\dagger}
{\tilde \phi}_{{\overline q}}^I\Bigr) \Bigr]
+\abs{\del_{\hmu}U_{R}(y)}^2\sum_{I}
\Bigl(\abs{{\tilde \phi}_{qI}}^2+
\abs{{\tilde \phi}_{{\overline q}}^I}^2\Bigr) \nonumber\\
\longrightarrow
&-&{{2\beta}\over L}{\rm tr}\bigl(\bar\psi_{1}\psi_{1}+ 
\bar\psi_{2}\psi_{2}\bigr)
+\bigl({\beta\over L}\bigr)^2\sum_{I}
\bigl(\abs{{\tilde \phi}_{qI}}^2 +\abs{{\tilde \phi}_{{\overline q}}^I}^2\bigr)
\nonumber \\ 
&+&{{2\beta g}\over L}\sum_{I}
\bigl({\tilde \phi}_{qI}^{\dagger}T^a
{\tilde \phi}_{qI}-{\tilde \phi}_{{\overline q}}^I
T^a{\tilde \phi}_{{\overline q}}^{I\dagger} \bigr)\Phi^a.
\label{eq:eac}
\end{eqnarray}
The $g$ is the gauge coupling constant.
The $\rightarrow$ in (\ref{eq:eac})
means that the dimensional reduction from $D=4$ to $D=3$ is carried 
out, ignoring the Kaluza-Klain modes $(n\ne 0)$ in the Fourier series. 
We have denoted the gauge field for the $S^1$-direction $A_{3}^a$ as $\Phi^a$. 
The $\Phi^a$ is real scalar field which belongs to the adjoint 
representation under the gauge group.
We realize again that
the supersymmetry breaking is entirely due to the locality of the 
gauged $U(1)_{R}$ transformation, {\it i.e.} $U_{R}\del_{\hmu}U_{R}^{\dagger}$ 
and $\abs{\del_{\hmu}U_{R}}^2$.
\par
We find that the supersymmetry breaking terms are the scalar mass 
and the trilinear scalar terms whose couplings depend only on an 
unique parameter $\beta$ and the gauge coupling $g$.
As the remarkable consequence, the supersymmetry breaking terms generated 
in this mechanism are common to all flavours and are soft breaking. 
This is because the derivative $\del_{\mu}$ are common to all flavours 
and has mass dimension one, so that the 
couplings generated through the derivative are always universal and 
dimensional couplings. 
The universality may be needed to avoid the FCNC.
\par
We also mention the supertrace of the squared mass matrix
${\rm Str}{\cal M}^2=
\sum_{J}(-)^{2J}(2J+1){\cal M}_{J}^2=0$, where $J$ stands for 
the spin of the particles.
In our case the masses for the gaugino and 
the quarks are $\beta/L$ and 
$m$, respectively. On the other hand, the mass for the squarks is 
$m^2-(\beta/L)^2$.
The gluon $A_{\mu}^a$ and the scalar $\Phi^a$ are massless. 
It is evaluated as\cite{fn}
\begin{equation}
{\rm Str}{\cal M}^2=2(m^2-(\beta/L)^2)-2(m^2+(\beta/L)^2)=-4(\beta/L)^2 \ne 0.
\label{eq:mstr}
\end{equation}
This result is expected because the supersymmetry is broken explicitly in 
our case. This may be desirable when we try to build models with the 
soft supersymmetry breaking terms based on our mechanism. 
\par
One may think that the boundary condition associated with the global 
flavour symmetry is possible. The boundary condition, however, does 
not break the supersymmetry because their charges of global flavour 
transformations are the same between the bosons and the fermions in a 
supermultiplet. In this case we would obtain supersymmetric invariant 
soft terms generated by the same manner discussed in this section.
\par
\section{Non-integrable Phase and Its Dynamics in SQCD}
We have discussed that the boundary condition associated with 
the $U(1)_{R}$ symmetry breaks the supersymmetry. In this 
section we shall discuss the role of the boundary condition
associated with the global gauge symmetry, which have been ignored in 
the previous section. 
Readers familiar with this topics can skip argument below and go to 
the subsection ${\bf 3.1}$ directly. 
Details discussions are given in\cite{ahosotani}.
\par
For general discussions we shall consider $SU(N)$ SQCD on 
$M^{D-n}\otimes T^n$ in this section, where $M^{D-n}$ and $T^n$ are
$(D-n)$-dimensional Minkowski space-time and $n$-torus, respectively. 
The $x^{\mu} (\mu=0, 1,\cdots, D-(n+1))$ are the coordinates of 
$M^{D-n}$, and $y^a (a=1,\cdots, n)$ are those of $T^n$. We use 
a notation $x^{\hmu}$ (or $x$) which stands for 
$x^{\hmu}\equiv x\equiv (x^{\mu}, y^a)$.
We define the boundary conditions as follows;
\begin{eqnarray}
A_{\hmu}(x^{\mu}, \cdots y^a+L^a \cdots )
=U_{g}^a A_{\hmu}(x^{\mu}, \cdots y^a \cdots)U_{g}^{a\dagger},\nonumber\\ 
\lambda(x^{\mu}, \cdots y^a+L^a \cdots ) 
={\rm e}^{i\beta}
U_{g}^a \lambda (x^{\mu}, \cdots y^a \cdots)U_{g}^{a\dagger}. 
\label{eq:boundaryb}
\end{eqnarray}
The $L^a (a=1, \cdots, n)$ is the length of the 
circumference of each circle, and 
a constant matrix $U_{g}^a (a=1\cdots n)\in SU(N)$. 
The lagrangian 
is still single-valued with the fields having the boundary conditions. 
We symbolically 
denote the boundary condition as $U_{g}^a (a=1, \cdots, n)$.
In order to fix a theory with the boundary conditions (\ref{eq:boundaryb}) 
definitely, a set of $U_{g}^a(a=1,\cdots, n)$ has given for any configurations 
of the fields $A_{\hmu}, \lambda$.
\par
There exists a class of gauge transformation which does not change 
the boundary condition $U_{g}^a$. We denote the class of the gauge 
transformation by $\Omega(x)$.
Under the gauge transformation $\Omega(x)$, the fields transform as
\begin{eqnarray*}
A_{\hmu}(x) \rightarrow
A_{\hmu}^{\prime}(x)&=&\Omega(x)A_{\hmu}(x)\Omega^{\dagger}(x)
+{i\over g}\Omega(x)\del_{\hmu}\Omega^{\dagger}(x) \\
\lambda(x)\rightarrow
\lambda^{\prime}(x)&=&\Omega(x)\lambda(x)\Omega^{\dagger}(x).\\
\end{eqnarray*}
The boundary conditions for the transformed 
fields $A_{\hmu}^{\prime}, \lambda^{\prime}$ are the same with those 
for $A_{\hmu}, \lambda$. Hence, we have
\begin{equation}
U_{g}^a=\Omega(x^{\mu}, y^a+L^a)U_{g}^{a}\Omega^{\dagger}(x^{\mu}, y^a).
\label{eq:bcsym}
\end{equation}
\par
We can consider another class of the gauge transformation by which 
the boundary condition is changed. Let us denote such gauge 
transformation by ${\cal T}(x)$. Under the transformation by ${\cal 
T}(x)$, the fields are redefined in the form of the gauge 
transformation by 
\begin{eqnarray*}
A_{\hmu}^{\prime}(x)&=&{\cal T}(x)A_{\hmu}(x){\cal T}^{\dagger}(x)
+{i\over g}{\cal T}(x)\del_{\hmu}{\cal T}^{\dagger}(x),\\
\lambda^{\prime}(x)&=&{\cal T}(x)\lambda(x) {\cal T}^{\dagger}(x).
\end{eqnarray*}
In order for the redefined fields $A_{\hmu}^{\prime}, 
\lambda^{\prime}$ to satisfy (\ref{eq:boundaryb}) with new boundary 
condition $U_{g}^{a\prime}$, we have
\begin{equation}
U_{g}^{a\prime}
={\cal T}(x^{\mu}, y^a+L^a)U_{g}^{a}{\cal T}^{\dagger}(x^{\mu}, y^a).
\label{eq:bcchange}
\end{equation} 
The gauge transformations ${\cal T}(x)$ ( and $\Omega(x)$ ) must satisfy 
$$
\del_{\hmu}\Bigl[{\cal T}(x^{\mu}, y^a+L^a)U_{g}^{a}{\cal T}^{\dagger}
(x^{\mu}, y^a)\Bigr]=0
$$
to maintain the $x$-independence of (new) boundary condition. 
\par
Let us assume $\vev{F_{\hmu\hnu}}=0$ in the vacuum, and it follows that
\begin{equation}
\vev{A_{\hmu}}={i\over g}V^{\dagger}(x)\del_{\hmu}V(x).
\label{eq:puregauge}
\end{equation}
It is important to note that this pure gauge 
configuration, in general, is physically distinct 
from $\vev{A_{\hmu}}=0$ in multiply-connected space because the 
transformation with ${\cal T}(x)$ change the boundary condition as seem 
from (\ref{eq:bcchange}). The $V(x)$ is determined by quantum effects 
as a function of $U_{g}^a$ up to a global gauge transformation.
If we perform the gauge transformation with ${\cal T}(x)=V(x)$, then 
we have $\vev{A_{\hmu}^{\prime}}=0$ with new 
boundary condition such as
\begin{equation}
U_{g}^{a:inv}\equiv U_{g}^{a\prime}=V(x^{\mu}, y^a+L^a)
U_{g}^{a}V^{\dagger}(x^{\mu}, y^a).
\label{eq:nipa}
\end{equation}
Since $U_{g}^{a:inv}$ is the boundary condition 
for $\vev{A_{\hmu}^{\prime}}=0$, the symmetry of the theory is 
generated by the generators of the gauge group 
which commute with $U_{g}^{a:inv}$.
The gauge transformation which satisfy 
\begin{equation}
\Omega(x^{\mu}, y^a+L^a)=U_{g}^{a:inv}\Omega(x^{\mu}, y^a)
U_{g}^{a:inv\dagger}(x^{\mu}, y^a)
\label{eq:bcsyma}
\end{equation}
is large gauge transformation under which the non-integrable phases 
are invariant. We will find that this invariance is reflected in the 
effective potential for the non-integrable phases.
\par
Let us show that $U_{g}^{a:inv}$ is closely related with the 
path-ordered integral along a loop $C$ of the compactified coordinate;
$$
W_{c}^a(C)\equiv {\cal P}{\rm exp}[-ig \int_{C}dy^a A_{a}]U_{g}, \quad 
(a=1,\cdots, n).
$$
We can define $n$ numbers of path-ordered integrals for each gauge 
field $A_{a} (a=1,\cdots, n)$ along each compactified 
coordinate $y^a (a=1,\cdots, n)$.
For the pure gauge configuration (\ref{eq:puregauge}), it is 
evaluated as
\begin{equation}
W_{c}^{a}(C)=V^{\dagger}(x^{\mu}, y^a)
V(x^{\mu}, y^a+L^a)U_{g}^{a}.
\label{eq:poia}
\end{equation}
Hence, from (\ref{eq:nipa}), $W_{c}^{a}(C)$ is related to $U_{g}^{a:inv}$ as
$$
W_{c}^{a}(C)=V^{\dagger}(x^{\mu}, y^a)U_{g}^{a:inv}V(x^{\mu}, y^a).
$$
The eigenvalues of $U_{g}^{a:inv}$ or $W_{c}^{a}(C)$ are called 
non-integrable phases, which are important quantities in discussing 
the gauge symmetry breaking of the theory.
\par
One of the boundary conditions $U_{g}^{a} (a=1, \cdots, n)$ for the 
gauge field $A_{a}$ can be 
diagonalized by utilizing the degrees of freedom of global gauge 
transformation. Here we assume that all of them have a diagonal form
as a special case;
\begin{equation}
U_{g}^a=
\left( \begin{array}{cccc}
{\rm e}^{i\lambda_{a,1}}&	& & \\
&\ddots & & \\
& &\ddots & \\
& & & {\rm e}^{i\lambda_{a,N}}
\end{array}\right),\quad \sum_{i=1}^N \lambda_{a, i}=0,\quad 
(a=1, \cdots, n),
\label{eq:hbc}
\end{equation}
where $i(=1, \cdots, N)$ stands for the indices of $SU(N)$ gauge group. 
Let us perform the gauge transformation ${\cal T}(y)$ given by 
\begin{equation}
{\cal T}(y)=
\left(\begin{array}{cccc}
{\rm e}^{i\varphi_{1}}& & & \\
& \ddots & & \\
& &\ddots & \\
& & & {\rm e}^{i\varphi_{N}}
\end{array}\right),\quad {\rm with}
\quad \varphi_{i}\equiv 
\sum_{b=1}^n{h_{b, i}\over L^b}y^b, 
\quad \sum_{i=1}^N h_{b, i}=0.
\label{eq:gtrf}
\end{equation}
Under the gauge transformation with (\ref{eq:gtrf}), the boundary 
condition (\ref{eq:hbc}) changes to
\begin{eqnarray}
\begin{array}{l}
U_{g}^{a\prime}={\cal T}(y^a+L^a)U_{g}^a{\cal T}^{\dagger}(y^a)\\ [2mm] = 
\left(\begin{array}{cccc}
{\rm e}^{i(\lambda_{a,1}+h_{a, 1})}& & \\ & \ddots & & \\
& & \ddots & \\
& & & {\rm e}^{i(\lambda_{a, N}+h_{a, N})} \end{array}\right).\\ [2mm]
\label{eq:mnbc}
\end{array}
\end{eqnarray}
We can redefine the fields so as to satisfy the periodic boundary 
condition by choosing $\lambda_{a, i}=-h_{a, i}$. 
New gauge field with $U_{g}^{a\prime}={\bf 1}_{N\times N}$ is 
redefined as
\begin{equation}
A_{\hmu}^{\prime}={\cal T} A_{\hmu}{\cal T}^{\dagger}+{i\over g} 
{\cal T}\del_{\hmu}{\cal T}^{\dagger}
={\cal T} A_{\hmu}{\cal T}^{\dagger}
+\delta_{\hmu, a}{1\over {gL^a}}
\left(\begin{array}{cccc}
h_{a, 1}& & & \\
& \ddots& & \\
& &\ddots & \\
& & & h_{a, N}
\end{array}\right).
\label{eq:redef}
\end{equation}
We see that the effect of the boundary condition shifts the 
gauge field by the constant, the second term in (\ref{eq:redef}). 
New gauge field $A_{a}^{\prime} (a=1,\cdots, n)$ satisfies the periodic 
boundary condition $U_{g}^{a\prime}={\bf 1}_{N\times N}$, so that 
we expect $\vev{A_{a}^{\prime}}$ to be constant.
Hence, we parameterize it as follows; 
\begin{eqnarray}
\begin{array}{l}
\vev{A_{\mu}^{\prime}} = 0,\quad (\mu =0, 1, \cdots D-(n+1)) \\[2mm] 
\vev{A_{a}^{\prime}} = {1\over{gL^a}}
\left (\begin{array}{cccc}
\theta_{a, 1}& & & \\
& \ddots & & \\
& & \ddots& \\
& & & \theta_{a, N}
\end{array}\right),\quad
\sum_{i=1}^{N}\theta_{a, i}=0. \\ [2mm]
\label{eq:expf}
\end{array}
\end{eqnarray}
The constant background gauge field (\ref{eq:expf}) can be written in the 
pure gauge form;
\begin{equation}
\vev{A_{a}^{\prime}}
={i\over g}V^{\prime\dagger}(y)\del_{a}V^{\prime}(y), 
\label{eq:npure}
\end{equation}
where $V^{\prime}(y)$ is given by
$$
V^{\prime}(y)=
\left(\begin{array}{cccc}
{\rm e}^{-i\sum_{b=1}^{n}{{\theta_{b, 1}}\over L^b}y^b} 
& & \\ & \ddots & 
& \\
& & \ddots&\\
& & & {\rm e}^{-i\sum_{b=1}^{n}
{{\theta_{b, N}}\over L^b}y^b}
\end{array}
\right)\in SU(N).
$$
Under the gauge transformation with $V^{\prime}(y)$,
new boundary condition for $\vev{A_{a}^{\prime\prime}}=0$ is obtained as
\begin{equation}
U_{g}^{a\prime\prime}=V^{\prime}(y^a+L^a)U_{g}^{\prime} 
V^{\prime\dagger}(y^a)
=\left(\begin{array}{cccc}
{\rm e}^{-i\theta_{a, 1}}& & & \\
& \ddots & & \\
& & \ddots & \\
& & & {\rm e}^{-i\theta_{a, N}}
\end{array}\right)\equiv U_{g}^{a:inv}.
\label{eq:bnip}
\end{equation}
On the other hand, the path-ordered integral for (\ref{eq:npure}) is 
evaluated as
$$
W_{c}^{a}(C)=V^{\prime\dagger}(x^{\mu}, y^a)
V^{\prime}(x^{\mu}, y^a+L^a)U_{g}^{a\prime}=
V^{\prime\dagger}(x^{\mu}, y^a)U_{g}^{a:inv}V^{\prime}(x^{\mu}, y^a),
$$
where we have used (\ref{eq:bnip}).
The $\theta_{a, i} (a=1, \cdots, n~;~i=1, \cdots, N)$ is the eigenvalue of 
$W_{c}^{a}$ or $U_{g}^{a:inv}$, that is, non-integrable phase.
\par
One can return to the original gauge field $\vev{A_{a}}$ 
from (\ref{eq:npure}) by the transformation with ${\cal T}^{-1}(y)$.  
We would also obtain 
the same $U_{g}^{a:inv}$ with (\ref{eq:bnip}) in this gauge.
\subsection{Effective Potential for Non-integrable Phases in SQCD}
Now, let us discuss the dynamics of the non-integrable phases, which
are dynamical degrees of freedom and can not 
be gauged away in a multiply-connected space.
We shall compute the effective potential for the non-integrable 
phases in the SQCD. 
It is worth noting that at
the tree-level the effective potential does not depend on the 
non-integrable phases, and the vacuum has the continuous 
degeneracy. The quantum effects, however, lift the degeneracy. We shall 
use the perturbation theory in one-loop approximation 
based on the background field method in the Feynman gauge.
The effective potential for the non-integrable phases have been 
evaluated in non-supersymmetric gauge theories
\cite{bhosotani}\cite{ahosotani}\cite{toms}\cite{higuchi}\cite{ho}.
\par
We understand that the effects of the boundary condition associated 
with the global gauge symmetry are interpreted as the constant background 
gauge field with the periodic boundary condition\cite{ftnta}
\begin{equation}
\vev{A_{a}}={1\over{gL^a}}
\left(\begin{array}{cccc}
\theta_{a, 1}& & & \\
& \ddots & & \\
& & \ddots & \\
& & & \theta_{a, N}
\end{array}
\right), (a=1, \cdots, n) \quad {\rm with}
\quad \sum_{i=1}^{N}\theta_{a, i}=0.
\label{eq:bgf}
\end{equation}
The $\theta_{a, i}$ is the eigenvalue of
the non-integrable phase which we shall determine dynamically. The field 
strength for (\ref{eq:bgf}) vanishes, $\vev {F_{\hmu\hnu}}=0$.
According to the prescription of the background field method, we 
obtain the effective potential for the non-integrable phases in the 
$D$-dimensional SQCD as follows;
\begin{eqnarray}
V_{eff}^{SQCD}&=&V_{g+gh}+V_{gaugino}\nonumber \\ 
&=&-{i\over 2}(D-2)\int_{k}{\rm 
Tr}~{\rm ln}~[D_{\hmu}D^{\hmu}] +{i\over 2}
(r 2^{[{D\over 2}]})\int_{k}{\rm Tr}~{\rm ln}~[D_{\hmu}D^{\hmu}].
\label{eq:effpot}
\end{eqnarray}
The $V_{g+gh}, V_{gaugino}$ stand for the contributions from gauge-ghost, 
gaugino, respectively. The $D-2$ in (\ref{eq:effpot}) is the number of 
on-shell bosonic degrees of freedom, and $r 2^{[D/2]}$ is the one of 
fermionic degrees of freedom, where
the factor $r=1/2$ if the gaugino $\lambda$ is 
a Majorana or Weyl, and $r=1/4$ if $\lambda$ is a Majorana-Weyl.
\par
We can expand the fields in the Fourier series for $T^n$-direction as
\begin{eqnarray}
\bigl[A_{\hmu}(x)\bigr]_{ij}
&=&{1\over{\sqrt{\prod_{a=1}^{n}L^a}}}
\sum_{n_{1}\cdots n_{n}}\bigl[A_{\hmu}^{(n_{1}\cdots
n_{n})}(x^{\mu})\bigr]_{ij} 
~{\rm exp}({\sum_{a=1}^{n}{{2\pi i}\over L^a}n_{a}y^a}), \nonumber \\
\bigl[{\lambda}(x)\bigr]_{ij}
&=&{1\over{\sqrt{\prod_{a=1}^{n}L^a}}}
\sum_{n_{1}\cdots n_{n}}\bigl[\lambda^{(n_{1}\cdots n_{n})}(x^{\mu})\bigr]_{ij} 
~{\rm exp}({\sum_{a=1}^{n}{{2\pi i}\over L^a} (n_{a}+{\beta\over 
{2\pi}})y^a}),
\label{eq:ftf}
\end{eqnarray}
where $i, j$ are indices of $SU(N)$ gauge group. The effect of the 
boundary condition associated with the $U(1)_{R}$ symmetry manifestly 
appear in the momentum of the compactified direction 
as
$p_{a}={{2\pi}\over L_{a}}(n_{a}+{\beta \over {2\pi}}) (a=1,\cdots, n)$
for the gaugino field. 
The covariant derivative $D_{\hmu}$ in (\ref{eq:effpot}) is the one with 
the constant background gauge field (\ref{eq:bgf}), which is evaluated as 
\begin{eqnarray*}
\Bigl[D_{\hmu}D^{\hmu}\Bigr]_{ij}&=&D_{\mu}D^{\mu}+D_{a}D^{a},\quad 
(\mu=0, 1,\cdots D-(n+1)~;~ a=1,\cdots, n)\\
&=&\del_{\mu}^2-D_{a}^2 \\
&=&\del_{\mu}^2-\sum_{a=1}^{n}
\bigl[\del_{a}-{i\over L^a}(\theta_{a, i}- \theta_{a, j})\bigr]^2. \\
\end{eqnarray*}
Then, (\ref{eq:effpot}) becomes
\begin{eqnarray}
V_{eff}^{SQCD}&=&{1\over 2}(D-2)\sum_{n_{1}\cdots n_{n}}\sum_{i, j=1}^{N} 
{1\over{\sqrt{\prod_{a=1}^{n}L^a}}} \nonumber \\ &\times&\int 
{{d^{D-n}p}\over {(2\pi)^{D-n}}} {\rm 
ln}~\Bigl[p^2+\sum_{a=1}^{n}\bigl({{2\pi}\over L^a}\bigr)^2 
\bigl(n_{a}-{{\theta_{a, i}-\theta_{a, j}}\over {2\pi}}\bigr)^2 
\Bigr]_{ij} \nonumber\\
&-&{1\over 2}r 2^{[D/2]}
\sum_{n_{1}\cdots n_{n}}\sum_{i, j=1}^{N} {1\over{\sqrt{\prod 
_{a=1}^{n}L^a}}} \nonumber \\ &\times& \int {{d^{D-n}p}\over 
{(2\pi)^{D-n}}} {\rm ln}~\Bigl[p^2+\sum_{a=1}^{n}\bigl({{2\pi}\over 
L^a}\bigr)^2 
\bigl(n_{a}-{{\theta_{a, i}-\theta_{a, j}-\beta}\over {2\pi}} 
\bigr)^2\Bigr]_{ij}.
\label{eq:effa}
\end{eqnarray}
The off-diagonal $(i\ne j)$ components contribute to the effective 
potential.
\par
It is clear that if $\beta=0$, the effective potential vanishes 
for any values of $\theta$'s in the 
space-time dimensions $D=3, 4, 6, 10$ because of the equal number of 
physical degrees of freedom between the bosons and the fermions in the 
supersymmetric Yang-Mills theory\cite{brink}; $$
(D-2)-r 2^{[D/2]}=0 \quad {\rm for}\quad \left\{
\begin{array}{ll}
D=3, & r={1\over 2},\\[2mm]
D=4, & r={1\over 2},\\[2mm]
D=6, & r={1\over 2},\\[2mm]
D=10, & r={1\over 4}.\\[2mm]
\end{array}
\right.
$$
The $\beta=0$ restores the supersymmetry to yield the vanishing effective 
potential for any values of $\theta$'s 
due to the non-renormalization theorem in the supersymmetric theories.
The eigenvalues of the non-integrable phases can not be determined 
dynamically in this case.
The boundary condition associated
with the $U(1)_{R}$ symmetry explicitly breaks the supersymmetry to yield 
non-vanishing effective potential. We can, in principle, determine 
the non-integrable phases dynamically as the minimum of  
the effective potential (\ref{eq:effa}).
\par
\subsubsection{Pure SQCD}
In order to demonstrate dynamical determination of the non-integrable 
phases as analytically as possible, we consider 
a toy model, $4$-dimensional SQCD 
compactified on $M^3\otimes S^1$. 
In this case the boundary conditions we take are reduced from 
(\ref{eq:boundaryb}) to
\begin{equation}
A_{\hmu}(x^{\mu}, y+L)=U_{g}A_{\hmu}(x^{\mu}, y)U_{g}^{\dagger}, \qquad 
\lambda(x^{\mu}, y+L)={\rm e}^{i\beta}U_{g}\lambda(x^{\mu}, 
y)U_{g}^{\dagger}. 
\label{eq:ymbc}
\end{equation}
The effective potential is evaluated as 
\begin{equation}
V_{eff}^{SQCD}=-(4-2){\Gamma(2)\over {\pi^2L^4}}\sum_{i, j=1}^{N} 
\sum_{n=1}^{\infty}{1\over n^4}\cos (n(\theta_{i}-\theta_{j})) \\ \nonumber
+{1\over 2}2^2{\Gamma(2)\over {\pi^2L^4}} \sum_{n=1}^{\infty}{1\over 
n^4}\cos (n(\theta_{i}-\theta_{j}-\beta)). \\ \nonumber
\label{eq:egeff}
\end{equation}
We keep only $\theta$-dependent terms after the momentum integrations 
in (\ref{eq:effa}).
Moreover, if we assume that the gauge group is $SU(2)$, we have only one 
order parameter $\theta_{1}=-\theta_{2}\equiv \theta$. The summation with 
respect to the Fourier mode $n$ can be done by the formula; 
\begin{equation}
F(t)\equiv \sum_{n=1}^{\infty}{1\over n^4}\cos nt =-{1\over 
{48}}t^2(t-2\pi)^2+
{{\pi^4}\over {90}}\quad {\rm for}\quad 0\le t \le 2\pi. \label{eq:fmr}
\end{equation}
Using this formula, we obtain
\begin{equation}
{\tilde V}_{eff}^{SQCD}\equiv {{V_{eff}^{SQCD}}\over {2/\pi^2L^4}}=-2F(t) 
+\bigl[F(t-\beta)+F(t+\beta)\bigr]
\equiv -2F(t)+G(t, \beta),
\label{eq:feff}
\end{equation}
where we have defined $t\equiv 2\theta$
and $G(t, \beta)\equiv F(t-\beta)+F(t+\beta)$. The $F(t)$ stands for the gauge 
and ghost contributions, and $G(t, \beta)$ stands for the gaugino 
contribution.
\par
The $F(t)$ and $G(t, \beta)$ have properties such 
that $F(t)=F(-t)=F(t+2\pi m)$ and $G(t, \beta)=G(-t, \beta)=G(t, -\beta) 
=G(t+2\pi m, \beta)=G(t, \beta+2\pi m)$, respectively. 
The $m$ is integer.
The effective potential has the periodicity $V_{eff}^{SQCD}(t, \beta)= 
V_{eff}^{SQCD}(t+2\pi m, \beta)$. The periodicity of the 
effective potential is traced back to the large gauge 
transformation generated by $\Omega (x)$ in (\ref{eq:bcsyma}). 
There is also the periodicity such as
$V_{eff}^{SQCD}(t, \beta)=V_{eff}^{SQCD}(t, \beta+2\pi m)$.
This periodicity follows from the redefinition of the gaugino field
$\lambda\rightarrow {\rm e}^{2\pi my/L}\lambda$.
\par
By straightforward calculations, we find 
$$
{\tilde V}_{eff}^{SQCD}=-{\beta^2\over {12}}\Bigl[ 3t^2-6\pi 
t+2\pi^2+\half\beta^2\Bigr]\ge {\tilde V}_{eff}(t=0~({\rm mod}~ 2\pi), 
\beta). 
$$
The eigenvalues of the non-integrable phases are determined dynamically 
at $\theta=0~({\rm mod}~\pi)$ as the minimum of the effective potential.
In this case the gauge symmetry is not broken because
$U_{g}^{inv}={\bf 1}_{2\times 2}$ commutes with all the generators of 
$SU(2)$ gauge group. 
Note that the parameter $\beta$ is not the order parameter of the 
effective potential.
\par
\subsubsection{Matters in the Fundamental Representation}
If we add massive matters $Q (\phi_{q}, q)$ and 
${\overline Q}(\phi_{\overline q}, {\overline q})$, then, 
additional contributions to the effective potential arise 
from quarks $q, {\overline q}$ and 
squarks $\phi_{q}, \phi_{\overline q}$. 
The $U(1)_{R}$ symmetry are defined by (\ref{eq:ursym}), from which we see
that the squarks have 
the $U(1)_{R}$ charges, but the quarks do not have them. The boundary 
conditions of these fields are defined by (\ref{eq:flbc}). 
\par
The effective 
potential for the non-integrable phases arising from the (s)quarks
is\cite{ftntb}
\begin{eqnarray}
V_{eff}^{fd}&=&V_{squarks}^{fd}+V_{quarks}^{fd}\nonumber \\ 
&=&-{i\over 2}(2N_{F}\times 2)\int_{k}{\rm Tr}~{\rm ln}~[D_{\hmu}D^{\hmu}] 
+{i\over 2}2N_{F}\times
(r 2^{[{D\over 2}]})\int_{k}{\rm Tr}~{\rm ln}~[D_{\hmu}D^{\hmu}], 
\label{eq:effpotm}
\end{eqnarray}
where $N_{F}$ is number of the flavour.
The superfields $Q, ({\overline Q})$ belong to the (anti) 
fundamental representation under $SU(2)$ gauge group. 
The covariant derivative $D_{\hmu}$ in (\ref{eq:effpotm}) is evaluated 
for these representations as 
\begin{eqnarray*}
\Bigl[D_{\hmu}D^{\hmu}\Bigr]_{ij}&=&D_{\mu}D^{\mu}+D_{a}D^{a},\quad (i=0, 
1,\cdots D-(n+1)~;~ a=1,\cdots n)\\ &=&\del_{\mu}^2-D_{a}^2 \\
&=&\Bigl[\del_{\mu}^2-\sum_{a=1}^{n}
\bigl[\del_{a}-{i\over L^a}\theta_{a, i}\bigr]^2 \Bigr]\delta_{ij}. \\
\end{eqnarray*}
Hence, we obtain
\begin{eqnarray*}
V_{eff}^{fd}&=&{1\over 2}2N_{F}\times 2\sum_{n_{1}\cdots 
n_{n}}\sum_{i=1}^{N}
{1\over{\sqrt{\prod_{a=1}^{n}L^a}}} \\
&\times&\int {{d^{D-n}p}\over {(2\pi)^{D-n}}} {\rm 
ln}~\Bigl[p^2+\sum_{a=1}^{n}\bigl({{2\pi}\over L^a}\bigr)^2 
\bigl(n_{a}-{{\theta_{a, i}-\beta}\over {2\pi}}\bigr)^2 \Bigr] \\ \nonumber
&-&{1\over 2}2N_{F}\times r 2^{[D/2]}
\sum_{n_{1}\cdots n_{n}}\sum_{i, j=1}^{N} {1\over{\sqrt{\prod 
_{a=1}^{n}L^a}}} \\
&\times& \int {{d^{D-n}p}\over {(2\pi)^{D-n}}} {\rm 
ln}~\Bigl[p^2+\sum_{a=1}^{n}\bigl({{2\pi}\over L^a}\bigr)^2 
\bigl(n_{a}-{{\theta_{a, i}}\over {2\pi}} \bigr)^2\Bigr].
\end{eqnarray*}
In the $4$-dimensions, this yields, keeping only $\theta$-dependent terms 
after the momentum integrations,
$$
V_{eff}^{fd}=
-2N_{F}\times 2{\Gamma(2)\over {\pi^2L^4}}\sum_{n=1}^{\infty} 
\sum_{i=1}^{N}{1\over n^4}\cos (n(\theta_{i}-\beta)) + 
2N_{F}\times \half 2^2{\Gamma(2)\over {\pi^2L^4}} 
\sum_{n=1}^{\infty}\sum_{i=1}^{N}{1\over n^4} \cos (n\theta_{i}),
$$
Then, for $SU(2)$ case, we obtain
\begin{eqnarray*}
{\tilde V}_{eff}^{fd}\equiv
{V_{eff}^{fd}\over {2/\pi^2L^4}}&=&N_{F}\Bigl[ -2\times 
\Bigl[F(t/2-\beta)+F(t/2+\beta)\Bigr] +2\times 2\times F(t/2) \Bigr]\\
&\equiv& N_{F}\Bigl[-2G(t/2, \beta)+
2\times 2\times F(t/2)\Bigr] \quad {\rm for}\quad 0\le t/2 \le 2\pi.
\end{eqnarray*}
The $G(t/2, \beta)\equiv F(t/2-\beta)+F(t/2+\beta), F(t/2)$ stands 
for the contributions from the squarks, quarks, respectively. 
As before, 
if $\beta=0$, the supersymmetry is restored, and 
${\tilde V}_{eff}^{fd}$ vanishes due to the non-renormalization theorem. 
\par
Adding these flavour contributions to the effective potential 
$V_{eff}^{SQCD}$, we obtain
\begin{eqnarray*}
{\tilde V}_{eff}^{SQCD+fd}&=&\Bigl[-2F(t)+G(t, \beta)\Bigr] 
+N_{F}\Bigl[-2G(t/2, \beta)+2\times 2\times F(t/2)\Bigr]\\
&=&{\beta^2\over {12}}\Bigl[
-3t^2+6\pi t-2\pi^2-\half\beta^2+N_{F}
\Bigl({3\over 2}t^2-6\pi t + 4\pi^2 +\beta^2\Bigr)
\Bigr].
\end{eqnarray*}
The $-2F(t)+G(t, \beta)$, which comes from the gluon and the gaugino, 
has minima at $\theta=0$ (mod $\pi$). On the other hand, 
$-2G(t/2, \beta)+2\times 2\times F(t/2)$, which comes from the 
(s)quark, has minima at $\theta=\pi$ (mod $2\pi$). 
Hence, ${\tilde V}_{eff}^{SQCD+fd}$
has minima at $\theta=\pi$ (mod $2\pi$) independent of $N_{F}$, so that
$U_{g}^{inv}=-{\bf 1}_{2\times 2}$.
The $SU(2)$ gauge symmetry is not broken for any numbers of the flavour
in the fundamental representation.
\par
\subsubsection{Matters in the Adjoint Representation}
Next, let us add massive matters in the adjoint representation under 
$SU(2)$ gauge group instead of those in the fundamental representation. 
We denote the superfield as $Q^{adj}$ which contain
adjoint-quark $q^{adj}_{I}$ and adjoint-squark $\phi_{qI}^{adj}$. 
The $U(1)_{R}$ 
symmetry is defined as the same way as the matters in the fundamental 
representation. We take the boundary conditions of the fields as
\begin{equation}
q_{I}^{adj}(x, y+L)
=U_{g}q_{I}^{adj}(x, y)U_{g}^{\dagger},\qquad
\phi_{qI}^{adj}(x, y+L)
={\rm e}^{i\beta}U_{g}\phi_{qI}^{adj}(x, y)U_{g}^{\dagger}. 
\label{eq:adbc}
\end{equation}
\par
The effective potential for the non-integrable phases arising from the 
adjoint-(s)quark is
\begin{eqnarray}
V_{eff}^{adj}&=&V_{squarks}^{adj}+V_{quarks}^{adj}\nonumber\\ 
&=&-{i\over 2}
(N_{F}^{adj}\times 2)\int_{k}{\rm Tr}~{\rm ln}~[D_{\hmu}D^{\hmu}] 
+{i\over 2}
N_{F}^{adj}\times (r 2^{[{D\over 2}]})
\int_{k}{\rm Tr}~{\rm ln}~[D_{\hmu}D^{\hmu}], 
\label{eq:effpotad}
\end{eqnarray}
The computation goes the same as before. We obtain
\begin{eqnarray*}
{\tilde V}_{eff}^{adj}\equiv
{V_{eff}^{adj}\over {2/\pi^2L^4}}&=&N_{F}^{adj}\Bigl[
-\Bigl[F(t-\beta)+F(t+\beta)\Bigr] +2F(t) \Bigr]\\
&\equiv& N_{F}^{adj}\Bigl[-G(t, \beta)+2F(t)\Bigr]\quad {\rm 
for}\quad 0\le t\le 2\pi.
\end{eqnarray*}
The $-G(t, \beta), 2F(t)$ stand for the squark, quark contributions, 
respectively. This result also corresponds to massless limit of 
the matters as before. The $\beta=0$ restores the supersymmetry to yield 
vanishing effective potential again. 
The total effective potential becomes
\begin{eqnarray*}
{\tilde V}_{eff}^{SQCD+adj}&=&\Bigl[-2F(t)+G(t, \beta)\Bigr] 
+N_{F}^{adj}\Bigl[-G(t, \beta)+2\times F(t)\Bigr]\\
&=&{\beta^2\over {12}}\Bigl[
-3t^2+6\pi t-2\pi^2-\half\beta^2+N_{F}^{adj}
\Bigl(3t^2-6\pi t + 2\pi^2 +\half\beta^2\Bigr)
\Bigr].
\end{eqnarray*}
For $N_{F}^{adj}\ne 1$, we find the minimum of the potential is located 
at $\theta=\pi/2$ independent of $N_{F}^{adj}$. This means 
$U_{g}^{inv}=-i\sigma^3$, so that the gauge symmetry is broken to 
$U(1)$. Even if we take $N_{F}^{adj}\rightarrow \infty$, the gauge 
symmetry is never restored to $SU(2)$. This is very contrary to the 
case for the matters in the fundamental representation.
\par
For $N_{F}^{adj}=1$, the effective potential vanishes for any values 
of $\theta$ even though $\beta$ is non-zero. We can not determine 
$\theta$ dynamically in this case. 
At this number of flavour, $W_{A}=(A_{\hmu}, \lambda)$ 
and $Q^{adj}=(\phi_{q}^{adj}, q^{adj})$ form a massless vector 
multiplet of $N=2$ super Yang-Mills theory in the $4$-dimensions. 
Note that the effective potential corresponds 
to massless limit of the matters. In $N=2$ supersymmetry there 
exists global $SU(2)_{R}$ symmetry in addition to 
the usual $U(1)_{R}$ symmetry.
Under $SU(2)_{R}$ transformation, $\lambda$ is 
interchanged into $q^{adj}$ and vice versa.
The $\lambda$ and $\phi_{q}^{adj}$ can form $N=1$ 
chiral superfield $Q^{adj}=(\lambda, \phi_{q}^{adj})$ 
instead of the original $N=1$ chiral superfield 
$Q^{adj}=(q^{adj}, \phi_{q}^{adj})$ by using $SU(2)_{R}$ symmetry. 
The $\lambda$ 
and $\phi_{q}^{adj}$ satisfy the same boundary condition as seen 
from (\ref{eq:ymbc}) and (\ref{eq:adbc}).
Hence, $N=1$ supersymmetry remains to be unbroken
though $N=2$ supersymmetry is broken by the boundary conditions. 
The non-renormalization theorem works by this unbroken $N=1$ 
supersymmetry, so that the 
effective potential vanishes even for non-zero values of $\beta$. 
\par
Finally, let us study the gauge symmetry breaking if we add
both matters in the fundamental representation 
and those in the adjoint representation.
The total effective potential is given by
\begin{eqnarray*}
{\bar V}_{eff}^{total}
&=&{\bar V}_{eff}^{SQCD}+{\bar V}_{eff}^{fd}+{\bar V}_{eff}^{adj}\\
&=&-2F(t)+G(t, \beta)
+N_{F}\Bigl[-2G(t/2, \beta)+4F(t/2)\Bigr]\\
&+&N_{F}^{adj}\Bigl[-G(t, \beta)+2F(t)\Bigr]\\
&=&{\beta^2\over {12}}\Bigl[-3t^3+6\pi t-2\pi^2-\half \beta^2
+N_{F}\Bigl({3\over 2}t^2-6\pi t +4\pi^2+\beta^2\Bigr)\\
&+&N_{F}^{adj}\Bigl(3t^2-6\pi t +2\pi^2+\half\beta^2\Bigr)
\Bigr].
\end{eqnarray*} 
For $(N_{F}, N_{F}^{adj})=(N_{F}, 0)~,(N_{F}, 1)$, 
$SU(2)$ gauge symmetry is not broken because $U_{g}^{inv}$ commutes 
with all the generators of $SU(2)$ in these cases. Note that we can not 
determine $\theta$ for $(N_{F}, N_{F}^{adj})=(0, 1)$ because of the 
vanishing effective potential due to the unbroken $N=1$ supersymmetry. 
Except for these cases the minimum of the total potential is 
located at 
$$
\theta={{N_{F}+N_{F}^{adj}-1}\over {N_{F}+2N_{F}^{adj}-2}}\pi,\quad
{\rm for}\quad 0\le \theta\le \pi.
$$
We obtain 
\begin{equation}
U_{g}^{inv}=\cos({{N_{F}+N_{F}^{adj}-1}\over{N_{F}+2N_{F}^{adj}-2}}\pi)
+i\sigma^3\sin({{N_{F}+N_{F}^{adj}-1}\over{N_{F}+2N_{F}^{adj}-2}}\pi),
\end{equation}
which commutes with only the third component of the $SU(2)$ 
generators $T^3$. Therefore, $SU(2)$ gauge symmetry is broken to $U(1)$.
\par
Let us study asymptotic behavior of the non-integrable phase
$\theta$. For fixed $N_{F}$, $\theta\rightarrow \pi/2$ as 
$N_{F}^{adj}\rightarrow \infty$, so that $U_{g}^{inv}\rightarrow 
-i\sigma^3$. The $SU(2)$ gauge symmetry is broken to $U(1)$ in this limit.
While, for fixed $N_{F}^{adj}$, $\theta\rightarrow \pi$ as $N_{F}\rightarrow 
\infty$, so that $U_{g}^{inv}=-{\bf 1}_{2\times 2}$. The $SU(2)$ 
gauge symmetry is restored in this limit. When 
$N_{F}=N_{F}^{adj}\rightarrow \infty$, $\theta$ approaches to
$2\pi/3$. The $U_{g}^{inv}$ takes the values of
$-\cos\pi/3-i\sigma^3\sin\pi/3$ in the limit. Hence, $SU(2)$ gauge 
symmetry is broken to $U(1)$ in this limit.
We depict how the gauge symmetry breaking depends on 
number and the representation under the gauge 
group of the matters in Figure $1$. 
\par
\subsubsection{Effective Lagrangian on $M^3$}
Taking into account of the discussions we have made, we obtain 
the effective lagrangian in the $3$-dimensions.
We begin from the $N=1$ SQCD in the $4$-dimensions and compactify 
it on $M^3\otimes S^1$ with the fields having the boundary 
conditions (\ref{eq:ymbc}) and (\ref{eq:flbc}) (or (\ref{eq:adbc})). 
Then, the effective lagrangian on $M^3$ takes the form of 
\begin{equation}
{\cal L}_{SQCD+matters}^{N=1}
\longrightarrow
{\cal L}_{SQCD+matters}^{eff}={\tilde {\cal L}}_{SQCD+matters}^{N=2}+
{\tilde {\cal L}}^{soft}. 
\label{eq:efflag}
\end{equation}
The ${\tilde {\cal L}}_{SQCD+matters}^{N=2}$ is the 
lagrangian of $N=2$ supersymmetric 
gauge theory with $SU(2)$ or $U(1)$ gauge group in the $3$-dimensions. 
In the $3, 4$-dimensional Minkowski 
space-time, the dimensions of the Dirac spinor are $2^{[D/2]}=2, 4$, 
respectively. From the point of view of the $3$-dimensions, $N=1$ 
supersymmetry in the original $4$-dimensions raises to $N=2$ 
supersymmetry in the $3$-dimensions by the dimensional reduction.
The $N=2$ supersymmetry in our effective theory consists of 
one vector multiplet and 
one (two) hyper multiplet(s), which are resulted from $N=1$ spinorial 
chiral superfield $W_{A}$ 
and $N=1$ chiral superfield $Q^{adj}(Q, {\overline Q})$, respectively 
by the dimensional reduction.
The $SU(2)$ gauge symmetry can be broken to $U(1)$ through the 
dynamics of the non-integrable phases, depending on number and the
representation under the gauge group of the matters. 
\par 
The ${\tilde {\cal L}}^{soft}$ stands for the soft
supersymmetry breaking terms due to the boundary condition associated with 
the $U(1)_{R}$ symmetry. Its explicit form is obtained in 
(\ref{eq:eac}).
If the matters belong to the adjoint representation under $SU(2)$ gauge group, 
$(T^a)_{bc}=if_{bac}$, while if they belong to the fundamental 
representation, $T^a=\sigma^a/2$, where $\sigma^a$ is the Pauli 
matrices. 
It should be understood that if the gauge symmetry is broken to 
$U(1)$, that is, $A_{3}^3=\Phi^3$ takes the non-zero 
vacuum expectation values, then,
physical field of $\Phi^a$ is the fluctuation
around the vacuum expectation value.
The effective lagrangian ${\cal L}_{SQCD+matters}^{eff}$ is the softly broken 
$N=2$ supersymmetric gauge theory. 
\par
\section{Conclusions and Discussions}
We have discussed the effects of the boundary conditions of the fields for the 
compactified directions on the supersymmetric gauge theory.
The supersymmetry can be broken explicitly by the 
boundary condition associated with the $U(1)_{R}$ symmetry. 
The effect of the breaking become manifest
by redefining the fields so as to satisfy the periodic boundary 
condition. They turn out to appear as the soft supersymmetry breaking terms
in the lagrangian.
The effects are always soft supersymmetry breaking 
because they are 
generated only through the derivative $\del_{\mu}$ in the kinetic terms, where 
the gauged $U(1)_{R}$ 
transformation by $U_{R}(y)$ is not respected as 
the symmetry of the theory. 
Remarkable feature is that the soft supersymmetry breaking terms do not have 
many arbitrary parameters, but they depend on an unique 
parameter $\beta$ and the gauge coupling. 
The soft supersymmetry breaking terms are common to 
all matters, which are needed to avoid the FCNC.
The supertrace of the squared mass matrix does not vanish because 
the supersymmetry is broken explicitly in our case.
It should be stressed that these desirable soft supersymmetry breaking 
terms are automatically incorporated into the theory by the boundary condition 
associated with the $U(1)_{R}$ symmetry. 
\par
We have also discussed the effects of the boundary condition 
associated with the global gauge symmetry.
The effects are interpreted as the constant background gauge 
field, which are dynamical degrees of freedom called non-integrable phases 
for the gauge field along the compactified direction in a multiply-connected 
space. Unlike the famous Aharanov-Bohm effects\cite{abohm},
the non-integrable phases are determined dynamically. 
We have shown explicitly that the gauge symmetry can be broken through the 
dynamics of the non-integrable phases in the SQCD.
The gauge symmetry breaking depends on numbers and the representation 
under the gauge group of the matters.
\par
We have begun from the $N=1$ SQCD in the $4$-dimensions and 
compactify it on $M^3\otimes S^1$ with the fields having  
the boundary conditions for the $S^1$-direction.
As the result of the effects of the boundary conditions, we have 
finally obtained the effective lagrangian on $M^3$, 
which is the softly broken $N=2$ supersymmetric gauge theory.
We can discuss both gauge and supersymmetry breaking  
in one scheme, say, the boundary conditions of the fields for 
the compactified directions. 
\par
We have evaluated the effective potential for the 
non-integrable phases (\ref{eq:effpotm}), (\ref{eq:effpotad}) in the massless 
limit of the matters.
The gauge symmetry breaking through the dynamics of non-integrable 
phases are essentially caused by the infrared dynamics of the theory.
The compactness of the extra coordinate $S^1$ shifts the zero point 
energies for massless particles, so that 
the gauge symmetry breaking is induced through the Casimir effect. 
If particles are massive, the gauge symmetry breaking 
may be different from the result obtained here\cite{chosotani}.
It is interesting to study how the massive particle affect the gauge 
symmetry breaking in the supersymmetric gauge theory.
We are studying the effects of the boundary conditions on more 
realistic higher dimensional supersymmetric gauge theories.
We believe 
that there are new possibilities for exploring models of softly 
broken supersymmetric gauge theories.
\vskip 2cm
\begin{center}
{\bf Acknowledgment}
\end{center}
\vspace{12pt}
The author would like to thank Professor C. S. Lim for fruitful 
discussions, encouraging me. This work was supported by Grant-in-Aid for 
Scientific Research Fellowship, No.5106.
\vspace{12pt}
\begin{center}
{\large{\bf Appendix}}
\end{center}
In this appendix we present explicit expressions which are omitted in 
the subsection ${\bf 2.3}$.
The $4$-dimensional lagrangian for the massive matters is given by
$$
{\cal L}_{matters}=\sum_{I}\Bigl[Q_{I}^{\dagger}{\rm e}^{2gV(N)}
Q_{I}+{\overline Q}^{I\dagger}{\rm e}^{2gV({\bar N})}{\overline Q}^{I}
\Bigr]_{D}+W(Q, {\overline Q})_{F},
$$
where the subscripts $F, D$ mean $F, D$-term, respectively, and  
$W(Q, {\overline Q})\equiv m_{I}{\overline Q}^IQ_{I}$.
The $g$ is the gauge coupling constant, and $I(=1,\cdots, N_{F})$ stands 
for flavour index.
The $V(N)\equiv V^aT^a, (V(\bar N)\equiv V^a{\overline T}^a)$ is 
the vector superfield in the Wess-Zumino gauge.
The $T^a(a=1,\cdots, N^2-1)$ is the generator of 
the gauge group, and ${\overline T}^a\equiv -T^{aT}=-T^{a*}$.
Under the supersymmetric transformations $\delta_{\xi}$ defined by
$$
\begin{array}{ll}
\delta_{\xi}\phi_{qI}={\sqrt 2}\xi q_{I},&
\delta_{\xi}\phi_{\overline q}^I={\sqrt 2}\xi {\overline q}^{I},\\[2mm]
\delta_{\xi}q_{I}={\sqrt 2}\xi F_{qI}+{\sqrt 2}i\sigma^{\hmu}{\bar\xi}
D_{\hmu}\phi_{qI},&
\delta_{\xi}{\overline q}^{I}={\sqrt 2}\xi F_{\overline q}^I
+{\sqrt 2}i\sigma^{\hmu}{\bar\xi}
D_{\hmu}\phi_{\overline q}^I,\\[2mm]
\delta_{\xi}F_{qI}
={\sqrt 2}i{\bar\xi}{\bar\sigma}^{\hmu}D_{\hmu}\phi_{qI}
+2g{\bar\xi}{\bar\lambda}\phi_{qI},&
\delta_{\xi}F_{\overline q}^I
={\sqrt 2}i{\bar\xi}{\bar\sigma}^{\hmu}D_{\hmu}\phi_{\overline q}^I
+2g{\bar\xi}{\bar\lambda}\phi_{\overline q}^{I},\\[2mm]
\end{array}
$$
the lagrangian varies as
${\cal L}_{matters}=\del_{\hmu}X^{\hmu}_{matters}$, where
$X^{\hmu}_{matters}\equiv X^{\hmu}_{Q}+X^{\hmu}_{\overline Q}$
is calculated as
\begin{eqnarray*}
X^{\hmu}_{Q}&=&\sum_{I}\Bigl[
ig\xi\sigma^{\hmu}{\bar\lambda}^a(\phi_{qI}^{\dagger}T^a\phi_{qI})
+{i\over{\sqrt 2}}\xi\sigma^{\hmu}{\bar q}_{I} F_{qI}
+{1\over {\sqrt 2}}\xi\sigma^{\hmu}{\bar\sigma}^{\hnu}q_{I}
(D_{\hnu}\phi_{qI})^{\dagger}\\
&+&{\sqrt 2}im_{I}
{\bar\xi}{\bar\sigma}^{\hmu}q_{I}\phi_{\overline q}^{I}\Bigr]
+h.c.
\end{eqnarray*}
and
\begin{eqnarray*}
X^{\hmu}_{\overline Q}&=&\sum_{I}\Bigl[
ig\xi\sigma^{\hmu}{\bar\lambda}^a(\phi_{\overline q}^{I\dagger}
{\overline T}^a\phi_{\overline q}^I)
+{i\over{\sqrt 2}}\xi\sigma^{\hmu}{\bar {\overline q}}^{I} F_{\overline q}^I
+{1\over {\sqrt 2}}\xi\sigma^{\hmu}{\bar\sigma}^{\hnu}{\overline q}^{I}
(D_{\hnu}\phi_{\overline q}^I)^{\dagger}\\
&+&{\sqrt 2}im_{I}
{\bar\xi}{\bar\sigma}^{\hmu}{\overline q}_{I}\phi_{qI}\Bigr]
+h.c.\quad .
\end{eqnarray*}
The lagrangian ${\cal L}_{matters}$
contains the kinetic terms for the quarks and the squarks. Their 
covariant derivatives are given by
$$
\begin{array}{ll}
D_{\hmu}\phi_{qI}
=\del_{\hmu}\phi_{qI}-igT^aA_{\hmu}^a\phi_{qI},&
D_{\hmu}q_{I}=\del_{\hmu}q_{I}-igT^aA_{\hmu}^aq_{I},\\[2mm]
D_{\hmu}\phi_{\overline q}^I
=\del_{\hmu}\phi_{\overline q}^I+ig\phi_{\overline q}^IT^aA_{\hmu}^a,&
D_{\hmu}{\overline q}^{I}=\del_{\hmu}{\overline q}^{I}
+ig{\overline q}^{I}T^aA_{\hmu}^a.\\[2mm]
\end{array}
$$
The derivative $\del_{\hmu}$ in the covariant derivatives 
do not respect the symmetry of the theory under the gauged $U(1)_{R}$ 
transformation by $U_{R}(y)={\rm e}^{i{\beta\over L}y}$.
The supersymmetry breaking terms are generated only through the derivative 
as discussed in the text.

\newpage 
\vspace{40pt}
\begin{center}
\leavevmode{\epsfxsize=12cm\epsffile{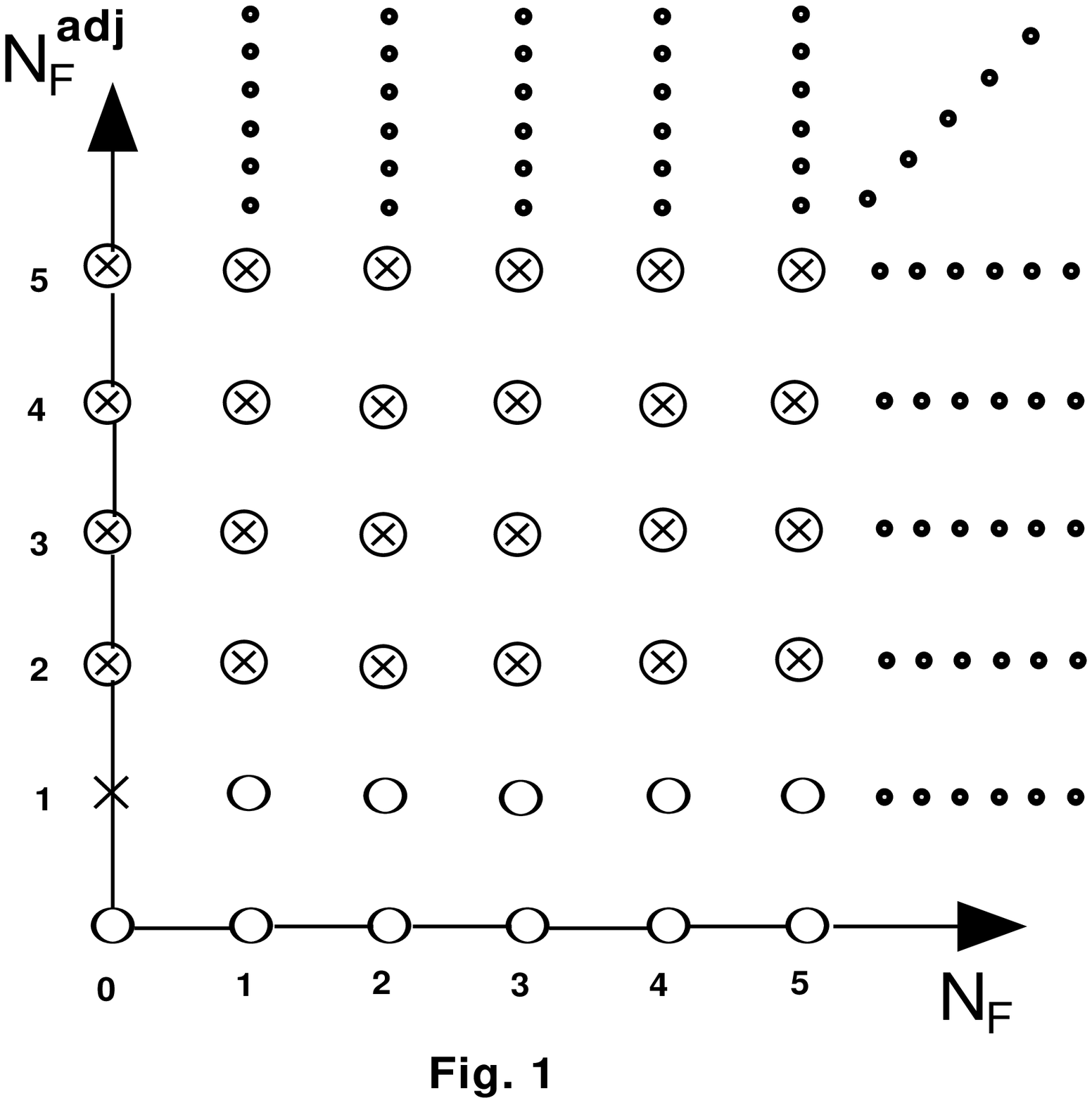}}
\end{center}
\smallskip
This figure shows how the gauge symmetry breaking depends on 
number and the representation under $SU(2)$ gauge group of the matters.
The $\bigotimes$ stands for the gauge symmetry breaking to $U(1)$. On the 
other hand, $\bigcirc$ stands for the unbroken $SU(2)$ symmetry.
At the point $(N_{F}, N_{F}^{adj})=(0, 1)$ denoted by $\times$ in the figure, 
we have unbroken $N=1$ supersymmetry. Details are discussed in the text. 

\begin{thebibliography}{99}
\bibitem{fn} In this evaluation we assume that the gauge symmetry is 
not broken. Even if the gauge symmetry is broken, ${\rm Str}{\cal 
M}^2$ does not vanish in our model.
\bibitem{ftnta} This form of the constant background gauge field 
follows from assuming the diagonal form of $U_{g}$. ( See (\ref{eq:hbc})). 
But we restrict ourself to flat connection $\vev{F_{\hmu\hnu}}=0$, so that 
the form of $\vev{A_{a}}$ on the torus is highly constrained such that all 
the components must commute. This yields the diagonal form of 
$\vev{A_{a}}$. On the other hand, if we assume the compactification 
on $M^{D-1}\otimes S^1$, we can 
always diagonalize $U_{g}$ as $U_{g}=C{\rm diag}({\rm e}^{i\lambda_{1}},\cdots 
{\rm e}^{i\lambda_{N}})C^{\dagger}$, where $C\in SU(N)$. In this case 
we can also write $\vev{A_{a}}$ as (\ref{eq:bgf}).
\bibitem{ftntb}This expression 
corresponds to the effective potential for massless (s)quarks. In this 
calculation we have taken the massless limit of the matters. The $U(1)_{R}$ 
symmetry is not exactly defined in the limit, but the similar 
non-trivial phases are allowed from the single-valueness of the 
lagrangian even if we begin from the massless (s)quarks.
\end{thebibliography}
\end{document}